\documentclass[10pt, a4paper, twocolumn]{article} 

\usepackage[svgnames]{xcolor}
\usepackage{amssymb}
\usepackage{pdfpages}
\usepackage{aasmacros}

\usepackage[english]{babel} 

\usepackage{microtype} 

\usepackage{amsmath,amsfonts,amsthm} 

\usepackage[svgnames]{xcolor} 

\usepackage[small, labelfont=bf, up, textfont=it]{caption} 

\usepackage{booktabs} 

\usepackage{lastpage} 

\usepackage{graphicx} 

\usepackage{enumitem} 
\setlist{noitemsep} 

\usepackage{sectsty} 
\allsectionsfont{\usefont{OT1}{phv}{b}{n}} 

\usepackage{geometry} 

\usepackage[T1]{fontenc} 
\usepackage[utf8]{inputenc} 

\usepackage{XCharter} 

\usepackage{fancyhdr} 
\renewcommand{\sectionmark}[1]{\markboth{#1}{}} 

\fancypagestyle{firstpage}{ 
	\fancyhf{}
}

\newcommand{\authorstyle}[1]{{\large\usefont{OT1}{phv}{b}{n}\color{DarkRed}#1}} 

\newcommand{\institution}[1]{{\footnotesize\usefont{OT1}{phv}{m}{sl}\color{Black}#1}} 

\usepackage{titling} 

\newcommand{\HorRule}{\color{DarkGoldenrod}\rule{\linewidth}{1pt}} 

\pretitle{
	\vspace{-15mm} 
	\HorRule\vspace{10pt} 
	\fontsize{24}{28}\usefont{OT1}{phv}{b}{n}\selectfont 
	\color{DarkRed} 
}

\posttitle{\par\vskip 15pt} 

\preauthor{} 

\postauthor{ 
	\vspace{1mm} 
	\par\HorRule 
	\vspace{2mm} 
}

\usepackage{lettrine} 
\usepackage{fix-cm}	

\usepackage{xstring} 

\usepackage[super,nomove]{cite}

\title{A D/H Ratio Consistent with Earth’s Water in Halley-type Comet 12P from ALMA HDO Mapping} 

\author{
	\authorstyle{Cordiner, M. A.$^{1,2}$, Gibb, E. L.$^{3}$, Kisiel, Z.$^{4}$, Roth, N. X.$^{1,5}$, Biver, N.$^{6}$, Bockel{\'e}e-Morvan, D.$^{6}$, Boissier, J.$^{7}$, Bonev, B. P.$^{5}$, Charnley, S. B.$^{1}$, Coulson, I. M.$^{8}$, Crovisier, J.$^{6}$, Drozdovskaya, M. N.$^{9}$, Furuya, K.$^{10}$, Jin, M.$^{1,2}$, Kuan, Y.-J.$^{11}$, Lippi, M.$^{12}$, Lis, D. C.$^{13}$, Milam, S. N.$^{1}$, Opitom, C.$^{14}$, Qi, C.$^{15}$, Remijan. A. J.$^{16}$} 
	\newline\newline 
	\textsuperscript{1}\institution{Solar System Exploration Division, NASA Goddard Space Flight Center, 8800 Greenbelt Road, Greenbelt, MD 20771, USA.}\\ 
	\textsuperscript{2}\institution{Department of Physics, Catholic University of America, Washington, DC 20064, USA.}\\ 
	\textsuperscript{3}\institution{Department of Mathematics, Physics, Astronomy, and Statistics, University of Missouri-St Louis, St Louis, MO, USA.}\\ 
	\textsuperscript{4}\institution{Institute of Physics, Polish Academy of Sciences, Al. Lotnik{\'o}w 32/46, 02-668 Warszawa, Poland.}\\ 
	\textsuperscript{5}\institution{Department of Physics, American University, 4400 Massachusetts Avenue, NW, Washington, DC 20016, USA.}\\ 
	\textsuperscript{6}\institution{LIRA, Observatoire de Paris, Universit{\'e} PSL, CNRS, Sorbonne Universit{\'e}, Universit{\'e} de Paris Cit{\'e}, 5 place Jules Janssen, 92195 Meudon, France.}\\ 
	\textsuperscript{7}\institution{IRAM, 300 Rue de la Piscine, 38406 Saint Martin d'Heres, France.}\\ 
	\textsuperscript{8}\institution{East Asian Observatory, 660 N. A’ohoku Place, Hilo, HI 96720, USA.}\\ 
	\textsuperscript{9}\institution{Physikalish-Meteorologisches Observatorium Davos und Weltstrahlungszentrum (PMOD/WRC), Dorfstrasse 33, 7260 Davos Dorf, Switzerland.}\\ 
	\textsuperscript{10}\institution{RIKEN Pioneering Research Institute, 2-1 Hirosawa, Wako-shi, Saitama 351-0198, Japan.}\\ 
	\textsuperscript{11}\institution{Center of Astronomy and Gravitation, and Department of Earth Sciences, National Taiwan Normal University, Taipei 116, Taiwan, ROC.}\\
	\textsuperscript{12}\institution{INAF – Osservatorio astrofisico di Arcetri, Largo Enrico Fermi 5, 50125 Firenze, Italy.}\\ 
	\textsuperscript{13}\institution{Jet Propulsion Laboratory, California Institute of Technology, 4800 Oak Grove Drive, Pasadena, CA 91109, USA.}\\
	\textsuperscript{14}\institution{Institute for Astronomy, University of Edinburgh, Royal Observatory, Edinburgh, EH9 3HJ, UK.}\\
	\textsuperscript{15}\institution{Institute for Astrophysical Research, Boston University, 725 Commonwealth Avenue, Boston, MA 02215, USA.}\\
	\textsuperscript{16}\institution{National Radio Astronomy Observatory, Charlottesville, VA 22903, USA.}
}

\date{Published in Nature Astronomy, 8 August 2025} 

\begin{document}

\maketitle 

\thispagestyle{firstpage} 


{\bf Isotopic measurements of Solar System bodies provide a primary paradigm within which to understand the origins and histories of planetary materials.  The D/H ratio in particular, helps reveal the relationship between (and heritage of) different H$_2$O reservoirs within the Solar System. Here we present interferometric maps of water (H$_2$O) and semiheavy water (HDO) in the gas-phase coma of a comet (Halley-type comet 12P/Pons-Brooks), obtained using the Atacama Large Millimeter/submillimeter Array (ALMA). The maps are consistent with outgassing of both H$_2$O and HDO directly from the nucleus, and imply a coma D/H ratio (for water) of $(1.71\pm0.44)\times10^{-4}$. This is at the lower end of the range of previously-observed values in comets, and is consistent with D/H in Earth's ocean water. Our results suggest a possible common heritage between a component of the Oort cloud's water ice reservoir, and the water that was delivered to the young Earth during the early history of the Solar System.}

{Due to their sensitivity to specific molecular formation pathways, as well as the local thermal and radiation environment, stable isotopic abundance ratios in solid or gas-phase molecules provide insight into the physical and chemical properties of the environment from which our Sun and planets emerged \cite{lyo05,fur17,vis18}}. As reviewed by ref. \cite{nom23}, the isotopic ratios of hydrogen, carbon, nitrogen and oxygen in comets indicate that a major component of cometary ice originated in cold ($\sim10$--30 K), dense gas during (or prior to) the formation of the Sun and planets. The combined effects of gas phase and grain-surface (photo-)chemistry resulted in strong (more than an order of magnitude) enrichment in deuterium, (factor of a few) enrichments in $^{15}$N, and more modest enrichments in $^{13}$C and $^{18}$O, in cometary ices. While some reprocessing of interstellar molecular isotopic ratios is expected to have occurred in the proto-Solar disk (and possibly, during the passage of interstellar matter into the disk), theoretical  models show that strong deuterium enrichment in the disk itself is likely to have been inefficient \cite{cle14,fur17}. Studying the D/H ratio in cometary H$_2$O (the dominant volatile species in comets) thus allows us to quantitatively investigate the link between the present day cometary water reservoir and the interstellar cloud from which our Solar System formed.

{Furthermore, the extent to which [D/H]$_{\rm H_2O}$ in the comet population may be consistent with Earth's oceans is of importance to theories regarding the origin of water, and therefore life, on our planet. Earth probably formed interior to the H$_2$O snowline in our protoplanetary disk, so the origin of its water and other volatiles remains a matter for debate \cite{mee20}. Likely sources include direct nebular gas accretion, inward-drifting icy pebbles, and later, collisions with ice-rich planetesimals, carbonaceous asteroids and comets. The D/H ratio provides one of several diagnostic measures for validating the possible contributions of these sources to the terrestrial volatile budget \cite{mar12a,ale18}.}

The first cometary D/H measurement was obtained by the Giotto spacecraft flyby of 1P/Halley, with a value of $(3.1\pm0.3)\times10^{-4}$ derived from water ion mass spectrometry \cite{ebe95b} (later revised to $(2.1\pm0.3)\times10^{-4}$ by \cite{bro12}). Measurements of [D/H]$_{\rm H_2O}$ have now been published for 11 different comets {(2 Halley-type comets (HTCs), 3 Jupiter-family comets (JFCs), and 6 Oort-cloud comets (OCCs))}, using various observational techniques, revealing a distribution in the range (1.4--6.5)$\times10^{-4}$ (or 0.9--4.2 times the Vienna standard mean ocean water (VSMOW) value of $1.56\times10^{-4}$) --- see \cite{nom23} and references therein. {Therefore, although comets were initially thought not to provide a good isotopic match for terrestrial water, this possibility is no longer formally excluded.}


While there appears to be some genuine diversity in [D/H]$_{\rm H_2O}$ across the comet population, most of the current statistics suffer from large error bars, and constitute a sensitivity-limited sample, intrinsically biased towards higher values of D/H. {Some of the scatter could be attributed to non-simultaneity of the HDO and H$_2$O observations (\emph{e.g.} 8P/Tuttle \cite{vil09}, C/1995 O1 \cite{mei98}, C/1996 B2 \cite{boc98}), a lack of reliable coma temperature data (\emph{e.g.} 46P \cite{lis19}), or the use of very different beam sizes to observe HDO and H$_2$O (\emph{e.g.} C/2014 Q2 and C/2012 F6 \cite{biv16})}. Improved statistics are required to confirm the true extent and shape of the distribution of cometary D/H values, as well as to better characterize any trends and differences in D/H between different dynamical classes of comets. Understanding the isotopic relationship between comets and other H$_2$O reservoirs throughout the Solar System (and beyond), thus remains an important work in progress, that the present study aims to address.  

12P/Pons-Brooks is a Halley-type comet on a highly elliptical 71-year orbit, last reaching perihelion at $r_H=0.78$~au on 2024 April 21. As a member of the class of nearly isotropic comets (NICs), 12P is likely to have come from the inner Oort-cloud \cite{lev96,nes17}, before gravitational interactions with the giant planets emplaced it into its current orbit. The NICs, which also include Oort cloud comets, likely formed in the vicinity of the giant planets, at a range of distances from the young Sun, before being scattered to the outer Solar System. Their chemical compositions are therefore expected to bear the imprint of the (range of) molecular and isotopic abundances prevalent in our protoplanetary disk midplane.

12P is among the brightest and most active known periodic comets, and thus presented an excellent opportunity for remote (ground-based) spectroscopy during its 2024 apparition. We used the ALMA telescope, as part of the Large Program 2023.1.01143.L, to obtain the first ever spatial maps of gas-phase H$_2$O and HDO (as well as other molecules) in an HTC. For details of the observations and data analysis see the Methods section.

The average, continuum-subtracted, nucleus-centered spectrum observed between 2024 April 18--25, in the spectral window covering the HDO $2_{1,1}$--$2_{1,2}$ line at 241.562 GHz, is shown in Extended Data Figure 1. HDO is clearly detected, in addition to multiple lines of methanol (CH$_3$OH) and ethylene glycol ((CH$_2$OH)$_2$, aGg' conformer).  Spectrally integrated maps of the observed H$_2$O and HDO emission are shown in Figure 1 (the H$_2$O $J_{K_a,K_c}=3_{1,3}$--$2_{2,0}$ line at 183.310 GHz was observed on 2024 May 4, separately from HDO).  CH$_3$OH maps are shown in Extended Data Figure 2.  All the molecular emission maps show a strong central peak, with evidence for diffuse emission at distances up to a few thousand kilometers away, consistent with quasi-isotropic expansion of gases from the sublimating nucleus. The inset emission line profiles show spectrally resolved velocity structure (with FWHM $\approx 1.7$~km\,s$^{-1}$), and evidence for a slight overall blueshift (between $\sim-0.1$ to $-0.2$~km\,s$^{-1}$) for all lines, consistent with more rapid outgassing in the direction of the observer, which can be interpreted as due to a higher gas outflow velocity on the warmer, day-side of the comet, considering the relatively low (Sun-target-observer) phase angle $\phi\sim30^{\circ}$. The enhanced strength of the redshifted CH$_3$OH peak relative to the blueshifted peak (particularly apparent in the higher-spectral-resolution May 4 data), can be explained by a lower expansion velocity on the night side of the comet, which results in a higher gas density. On the other hand, the greater optical thickness of H$_2$O causes suppression of the line flux from the more distant (redshifted) hemisphere of the coma, with enhanced, blueshifted H$_2$O emission from the nearside.

Due to noise, optical depth, and molecular excitation effects, the HDO/H$_2$O ratio cannot be directly inferred from the ALMA maps. Detailed radiative transfer modeling is needed for further analysis.

\begin{figure*}
\centering
\includegraphics[width=\textwidth]{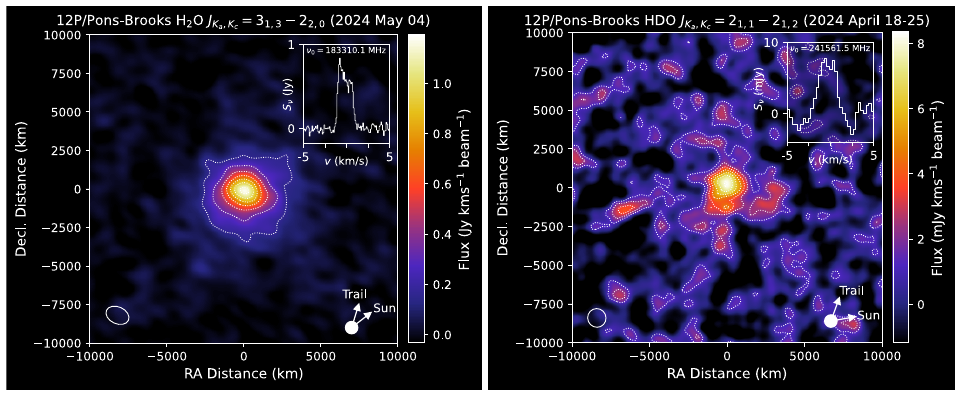}
\caption{{\bf ALMA emission maps of H$_2$O and HDO in comet 12P/Pons-Brooks.} The image cubes have been spectrally integrated, continuum-subtracted, and are centered on the thermal continuum peak. The contour separation is $5\sigma$ for H$_2$O, and $1\sigma$ for HDO, where $\sigma$ is the RMS noise of each map. Inset panels (upper right) show the spectral line flux ($S_{\nu}$) profiles extracted at the origin, as a function of cometocentric velocity ($v$). The spatial resolution (beam FWHM) is shown lower left, and the directions of the (sky-projected) comet-Sun and orbital trail vectors are shown lower right. The coordinate axes are sky-projected offsets (in kilometers) from the continuum peak, aligned with the celestial RA (right ascension) and Dec. (declination) grid. 
\label{fig:maps}}
\end{figure*}

Following the methods of Cordiner et al. \cite{cor23}, the calibrated, continuum-subtracted ALMA data were modeled directly (\emph{i.e.}, prior to imaging) using a spectrally resolved, Fourier-domain least-squares fitting approach. Since H$_2$O and HDO were observed approximately 2 weeks apart, and the comet's water production rate observed by SOHO/SWAN was variable over this period (M. Combi, private communication 2024), we used CH$_3$OH as a proxy for the water production rate appropriate for our HDO observation dates (2024 April 18--25), based on the measured CH$_3$OH/H$_2$O production rate ratio on 2024 May 4. We emphasize that this approach is applicable because (1) HDO and CH$_3$OH were always measured simultaneously (\emph{e.g.} Extended Data Figure 1), and (2) the CH$_3$OH/H$_2$O mixing ratio (derived from independent observations using the NASA Infrared Telescope Facility; IRTF), did not vary significantly during the period of our ALMA observations (see Extended Data Table 1).  Multiple lines of CH$_3$OH were observed on all dates, and were used to determine the (radially dependent) coma temperature profile (see Extended Data Figure 3).

Incorporating this temperature profile into our radiative transfer model, we subsequently used a 3D coma model (with hemispherical asymmetry along the Sun-comet line), to retrieve the total H$_2$O production rate on May 4 ($Q({\rm H_2O})=(4.71\pm0.09)\times10^{29}$~s$^{-1}$), leading to an average HDO/H$_2$O mixing ratio between April 18--25 of $(3.41\pm0.89)\times10^{-4}$ (where the uncertainty includes the combined sources of error outlined in the Methods section). Best-fit parameters and their uncertainties for the hemispherically-asymmetric HDO, H$_2$O and CH$_3$OH models are shown in Table \ref{tab:results}, including production rates ($Q$), outflow velocities in the sunward ($v_1$) and anti-sunward ($v_2$) directions, and the ratio of sunward-to-anti-sunward hemispheric production rates ($Q_1/Q_2$). {The multi-line IRTF H$_2$O data show ortho-to-para ratios (OPRs) consistent with the equilibrium statistical value of 3 (ref. \cite{ham16}) on all dates, so this value was used to convert the ALMA p-H$_2$O production rate to a total $Q({\rm H_2O})$.}  Due to the H-atom number statistics, our retrieved HDO/H$_2$O ratio corresponds to [D/H]$_{\rm H_2O}$ = $(1.71\pm0.44)\times10^{-4}$. This is consistent with the terrestrial (VSMOW) value of $1.56\times10^{-4}$, within the uncertainties.

\begin{table*}
\caption{Production rates and abundances of HDO, CH$_3$OH, and H$_2$O in the coma of 12P/Pons-Brooks\label{tab:results}}
\begin{center}
{\small
\vspace*{-3mm}
\begin{tabular}{lccccccc}
\hline
\hline
Molecule&Date(s)&$Q^a$&$Q_1/Q_2$$^b$&Abundance$^c$ & $v_1$ & $v_2$ & $\chi^2_R$$^d$\\
        & (UT)  & ($10^{26}$\,s$^{-1}$) & & (\%) & (km\,s$^{-1}$)& (km\,s$^{-1}$)&\\
\hline
HDO     & 2024 Apr 18--25&$3.15\pm0.42$&$1.7\pm0.5$&$0.034\pm0.006$&$1.12\pm0.12$ & $0.69\pm0.10$&1.03\\
CH$_3$OH& 2024 Apr 18--25&$214.8\pm0.3$&$1.133\pm0.003$&2.328$^e$$\pm0.003$&$0.907\pm0.001$&$0.654\pm0.001$&1.08\\
p-H$_2$O  & 2024 May 4 &$1178\pm22$&$1.56\pm0.07$&0.25$^f$&$0.90\pm0.01$&$0.75\pm0.02$&0.98\\
CH$_3$OH& 2024 May 4   & $109.6\pm0.8$&$1.15\pm0.18$&$2.328\pm0.016$&$0.893\pm0.005$&$0.655\pm0.003$&0.98\\
\hline
\end{tabular}
}
\parbox{0.96\textwidth}{\footnotesize
\vspace*{1mm}
$^a$ Total production rate.
$^b$ Ratio of sunward (day-side) to antisunward (night-side) production rates.
$^c$ Mixing ratio relative to H$_2$O.
$^d$ Reduced chi-square statistic.
$^e$ The $Q({\rm H_2O})$ value in our model was varied until the best-fit CH$_3$OH abundance matched that derived for May 4.  
$^f$ The p-H$_2$O/H$_2$O mixing ratio was fixed at 0.25 (see Methods). The $\pm$ error ranges are statistical ($1\sigma$) uncertainties from the SUBLIME fit covariance matrix.
}
\end{center}
\end{table*}

The real parts of the visibility fluxes as a function of antenna separation ($uv$ distance) are shown for H$_2$O and HDO in Figure 2, with best-fitting model curves overlaid. The large number of visibility data points (15786 visibilities $\times35$ spectral channels for HDO, and 1406 visibilities $\times70$ spectral channels for H$_2$O) necessitate additional averaging for display and interpretation purposes. Therefore in these plots, the observed fluxes have been combined into 10-m wide $uv$ bins, with additional averaging across the full-widths of the spectral lines (errors propagated accordingly). As shown by \cite{cor14,rot21,cor23}, such visibility plots are sensitive to the spatial distributions of the observed molecules, allowing nucleus (parent) \emph{vs.} coma (daughter) production mechanisms to be readily distinguished. For H$_2$O, HDO and CH$_3$OH, a parent model was found to provide an excellent fit to the observations; the quality of the fits (as monitored using the $\chi^2$ statistic) did not significantly improve through addition of distributed (coma) sources for any of the three molecules. To further visualize the quality of these model fits, the H$_2$O and HDO visibility spectra are shown in Extended Data Figure 4, averaged between $uv=0$--200 m. Since the CH$_3$OH lines are stronger (and therefore have higher signal-to-noise), their spectra and associated model fits can be usefully visualized as a function of antenna baseline across multiple $uv$ bins (as shown in Extended Data Figures 5 and 6).

We performed a $\chi^2$ analysis to investigate the origin of H$_2$O and HDO gas in the coma of comet 12P, by calculating the impact of a varying production scale length ($L_p$) on the goodness-of-fit (see Methods). The ALMA H$_2$O and HDO visibility data are found to be most consistent with nucleus sublimation ($L_p=0$). For H$_2$O and HDO, the 99\% confidence upper limits on their respective $L_p$ values are 135~km and 610~km; the latter being larger due to the lower signal-to-noise of the HDO data.  In reality, H$_2$O gas in the coma could originate from a mixture of nucleus and icy grain sublimation sources (\emph{e.g.} \cite{lis19}). Assuming 50\% of the H$_2$O is from a distributed (icy grain) source, the ALMA data are compatible with production at larger scale lengths $L_p<237$~km (99\% confidence), whereas in this case, the HDO $L_p$ value is unconstrained as a result of the lower signal-to-noise. Due to the combination of the spatial coverage of the interferometer and the distribution of the HDO emission, our ALMA data are most sensitive to D/H from coma gases at radial distances $\sim600$--6000 km from the nucleus --- \emph{i.e.}, outside the range of distances where icy grain sublimation of H$_2$O is likely to have been significant. Our measured D/H ratio is therefore not expected to depend on fractionation effects related to icy grains, which were proposed by \cite{mandt24} to explain radial variation of the HDO/H$_2$O ratio in comet 67P. {In addition, chemical-hydrodynamical coma models \cite{rod02} have shown that the HDO/H$_2$O ratio is not significantly affected by gas-phase fractionation, and is therefore representative of the nucleus ratio}.

\begin{figure*}
\centering
\includegraphics[width=\textwidth]{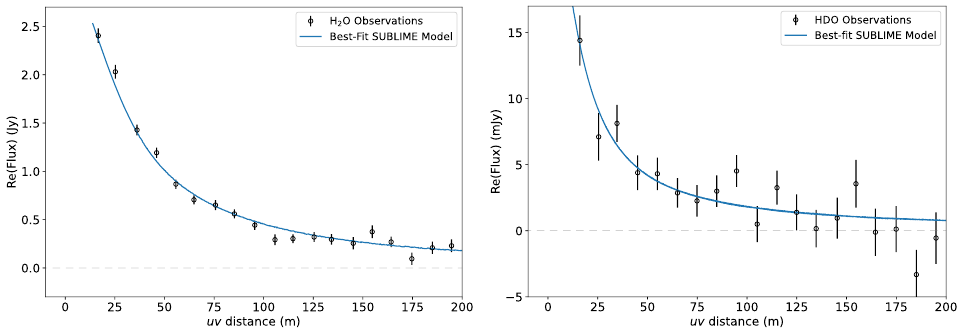}
\caption{{\bf Interferometric data for H$_2$O and HDO in comet 12P.} Binned real visibility fluxes are plotted for H$_2$O (left) and HDO (right) as a function of antenna baseline ($uv$ distance). Best-fit SUBLIME radiative transfer model curves are overlaid in blue, showing an excellent fit for both species assuming outgassing directly from the nucleus (no production in the coma). Error bars are $\pm1\sigma$ statistical uncertainties, with respect to the mean flux in each bin.
\label{fig:vis}}
\end{figure*}

Our D/H ratio for comet 12P is shown for comparison with other Solar System measurements (including all cometary measurements to date) in Figure 3. Combining our new D/H value with the previous measurements, including the revised values for comets 1P/Halley \cite{bro12}, C/1996 B2 \cite{lis19} and 67P/Churyumov-Gerasimenko \cite{mandt24}, the average D/H in cometary water is now $2.9\times10^{-4}$ (with a standard deviation of $1.4\times10^{-4}$), which is equivalent to $1.9\pm0.9$ times VSMOW. Our value in comet 12P therefore lies towards the lower end of the range of D/H previously found in all comets, and is consistent with the near-VSMOW values observed in the JFCs 46P/Wirtanen \cite{lis19} and 103P/Hartley 2 (ref. \cite{har11a}), and the OCC C/2014 Q2 (Lovejoy) \cite{biv16}. {The D/H ratio in 12P/Pons-Brooks is also within $1\sigma$ of the value for comet 1P/Halley \cite{bro12} and within $1.7\sigma$ of the value for 8P/Tuttle \cite{vil09}. While these results suggest that Halley-type comets might share a common D/H ratio, more observations are needed to confirm this.} To help determine the extent to which the 12P nucleus composition may be representative of the broader class of HTCs (and the comet population in general), a more complete investigation of its chemical taxonomy will be performed in a future article, based on the molecular abundances measured using our complete ALMA Large Program dataset, combined with observations from other telescopes. 

The distribution of D/H values in HTCs can be compared with that of JFCs and Oort cloud comets (OCCs) using the Kolmogorov-Smirnov (K-S) test \cite{mas51}, to search for statistical differences between these three classes of comet. Based on the data in Figure 3, we find the K-S test $p$ value --- \emph{i.e.} the probability that the observed differences between the HTC and OCC D/H measurements could occur, if they originated from the same underlying distribution --- is 99.7\%. The $p$ value for HTCs compared with JFCs is 77\%, and for JFCs compared with all NICs, $p=58$\%. Therefore, although differences between these cometary families cannot be ruled out based on the (presently limited) statistics, we find no significant reason to distinguish them based solely on D/H ratios. Based on chemical models of the Solar protoplanetary disk for which D/H varies as a function of radius \cite{wil07,alb14,fur17}, this implies that the different cometary families could have all accreted their ices over a similar range of distances {(within a few 10's of astronomical units of the proto-Sun)}.  

\begin{figure*}
\centering
\includegraphics[width=0.8\textwidth]{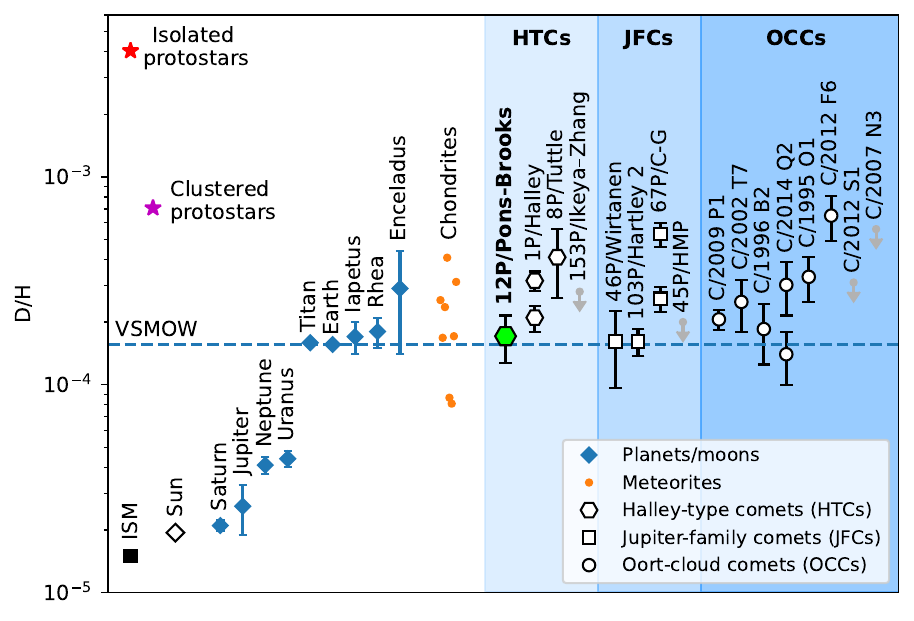}
\caption{{\bf Comparison of D/H ratios in different bodies.} D/H in water for comet 12P/Pons-Brooks is shown, along with previously-observed comets and other Solar System bodies, protostars, and the bulk interstellar medium (ISM) (see ref. \cite{nom23} and references therein). The recently revised value for comet 67P/Churyumov-Gerasimenko \cite{mandt24} has been added, along with the updated value for C/1996 B2 \cite{lis19}, and the two values derived by different authors for 1P/Halley \cite{ebe95b,bro12}. Upper limits are shown with grey arrows. The ISM and Sun values are for atomic hydrogen, giant planet values are for H$_2$, and Titan's value is for CH$_4$ (from \cite{nix12}). All other values are for H$_2$O. Error bars take into account various sources of random and systematic uncertainty appropriate for each method (see Methods section, and ref. \cite{nom23}, and references therein, for details).
\label{fig:dh}}
\end{figure*}

Our result confirms the previously identified $\sim1$ order of magnitude enrichment in deuterium of cometary water compared with the Solar and (bulk) interstellar elemental values \cite{cec14,nom23}. The D/H value in comet 12P is therefore consistent with preferential incorporation of D-atoms into water ice during the low-temperature evolutionary stage of the Sun’s natal molecular cloud, which is expected to occur prior to gravitational collapse and protoplanetary disk formation, but after the cloud is sufficiently dense to be shielded from interstellar radiation \cite{taq14,fur17}. However, the relatively modest [D/H]$_{\rm H_2O}$ enrichment compared to the $\sim$ 2 orders of magnitude found in some nearby protostellar envelopes \cite{nom23} implies that some re-equilibration of the cometary D/H ratio may have occurred, at the warmer temperatures ($\sim$ a few hundred to a few tens of Kelvin at $\sim1$--10 au radii \cite{alb14,eis18}), expected to have been present in the inner planet-forming regions of our protoplanetary disk midplane. Radial (or vertical) mixing and subsequent freeze-out of water vapour with a low D/H ratio, more equilibrated with high-temperature hydrogen in the disk, interior to the H$_2$O snow-line, could also have resulted in a reduced cometary D/H ratio.   According to the 2D turbulent-disk chemical model of \cite{alb14}, D/H values similar to that found in comet 12P ($\sim$ VSMOW) can be attained in the midplane ices at radial distances $\lesssim10$~au from the proto-Sun (\emph{i.e.}, in the vicinity of, or interior to, where Jupiter and Saturn formed \cite{tsi05}). 

Alternatively, a relatively warm environment in the Solar prestellar core may be implied, perhaps as a result of irradiation by other nearby stars (or supernovae) in the Sun's proto-cluster, thus inhibiting the fractionation of D into H$_2$O in interstellar ices, which primarily occurs at gas {and dust} temperatures below 30~K \cite{cec14,nom23}. {In the gas phase, deuterium enrichment is theorized to originate from H--D atom exchange in the reaction H$_3^+$ + HD $\rightleftarrows$ H$_2$D$^+$ + H$_2$, which proceeds more efficiently in the forward direction at the low ($\sim10$~K) temperatures found in dense interstellar clouds \cite{mil89}, leading to a reservoir of reactive deuterium that is subsequently incorporated into H$_2$O and other molecules \cite{taq14}. At higher ($\gtrsim 30$~K) temperatures, this deuteration pathway is suppressed. However, as shown by \cite{mil89}, deuterated organic molecules (such as DCN, HDCO, and CH$_3$D) can still be efficiently produced at these higher temperatures, starting from the gas-phase reaction CH$_3$$^+$ + HD $\longrightarrow$ CH$_2$D$^+$ + H$_2$. The significantly elevated D/H ratios found in cometary organics ($\sim2\times10^{-3}$ for HCN and CH$_4$ in C/1995 O1 and 67P, respectively \cite{mei98b,mul22}), could therefore also be consistent with the warm Solar prestellar core scenario.}

{Diffusive ice-mantle chemistry at warm (20--40~K) interstellar cloud temperatures has been theorized to result in enhanced abundances of complex organic molecules such as methyl formate, dimethyl ether, glycolaldehyde and ethylene glycol \cite{gar08}. While the large abundances of glycolaldehyde and ethylene glycol in comet C/2014 Q2 \cite{biv15} could be consistent with such a phenomenon, the non-detection of dimethyl ether in any comet to-date is at odds with this idea.  Additional measurements of the abundances of complex organic molecules in comets (including in 12P) would help clarify the situation. Ultimately, improved disk models incorporating detailed chemistry as well as radial mixing of gas, dust, pebbles and planetesimals, and infall from the interstellar medium, will be required for a more complete interpretation of the observed cometary D/H values.}


{In summary, the [D/H]$_{\rm H_2O}$ ratio in comet 12P/Pons-Brooks is consistent with terrestrial ocean water, supporting the idea that cometary impacts may have delivered water and other volatiles to Earth (including the biogenic elements, CHNOPS)}. We therefore reaffirm the importance of cometary compositional studies as part of a continued effort to understand the inventory of volatiles available on planetary surfaces during the early history of the Solar System, that may have been important for the origin of life. It will be important to expand our D/H statistics to include more comets, to determine whether or not the Halley-type comets share a common value, and to better characterize the distribution of cometary D/H ratios in general. {Additional D/H measurements in comets, including NICs and JFCs, would thus improve our understanding of how water was processed and isotopically fractionated during Solar System formation.}

\section*{Methods}

\subsection*{ALMA Observations.}

ALMA observations of H$_2$O, HDO and CH$_3$OH in comet 12P/Pons-Brooks were performed as part of program 2023.1.01143.L (PI: M. Cordiner), spread over 10 observing sessions (6 dates) between UT 2024 April 18 and 2024 May 4. Details of the observing geometry for each session are given in Supplementary Table 1. For the first 9 sessions (between 2024 April 18 and 2024 April 25), the correlator was configured to simultaneously observe HDO ($J_{K_a,K_c}=2_{1,1}$--$2_{1,2}$) at 241.561 GHz and CH$_3$OH ($J_K=5_K$--$4_K$ band) near 242 GHz in Band 6, with a spectral resolution of 488~kHz. For the remaining session on 2024 May 4, p-H$_2$O ($J_{K_a,K_c}=3_{1,3}$--$2_{2,0}$) at 183.310 GHz and CH$_3$OH ($J_K=4_K-3_K$ band) at 193~GHz were observed simultaneously in Band 5, at 122 kHz resolution. There were between 41--46 active antennas in the array, spanning baselines 15--500~m (corresponding to ALMA configurations C-2 to C-3). An aggregate continuum bandwidth of $\sim4$~GHz was also observed simultaneously with the spectral lines in each setup. Weather conditions were nominal for the observed frequency bands throughout all observations, with zenith precipitable water vapour (PWV) in the range 0.5--1.8~mm.  For further details of the individual spectral lines detected, see Supplementary Table 2. 

The position of the comet on the sky was tracked in real-time using the latest available ephemeris solution from the JPL Horizons database on each date. Data were flagged and calibrated in CASA version 6.5 (Ref. \cite{casa22}) using automated pipeline calibration scripts supplied by the Joint ALMA Observatory (JAO). Calibration of the amplitudes and bandpasses was performed using observations of the nearby quasar J0423-0120, while quasars J0407+0742, J0401+0413, J0424+0036 and J0309+1029 were used for phase calibration (selected dynamically for each session). For all observations, the comet was found to be offset by 1.7$''$--2.2$''$ north of the predicted ephemeris position. To facilitate subsequent data analysis, this offset was corrected in each measurement set, by fitting a point-source to the continuum visibility data and using the CASA {\tt phaseshift} task to move the fitted flux peak to the phase center. The thermal continuum was then subtracted from the visibilities using the {\tt uvcontsub} task, with a 2nd-order polynomial fit, excluding channels containing spectral line emission. The spectral axis of the visibility data (as a function of time) was corrected to the rest frame of the comet using the {\tt cvel2} task, then averaged across pairs of adjacent channels to reduce the frequency correlation introduced as a result of Hanning smoothing by the ALMA correlator. The 9 sets of Band 6 observations were then time-averaged and concatenated together to form a single measurement set.  

Imaging and deconvolution was performed using the {\tt tclean} (H{\"o}gbom) algorithm, with natural weighting, a flux threshold of twice the RMS noise per channel ($\sigma_c$) in each image, and a $10''$-diameter circular mask about the phase center. The resulting Band 6 image cube (covering HDO and CH$_3$OH) has a restoring beam FWHM of $1.10''\times1.03''$, with $\sigma_c=0.68$~mJy at 242~GHz. In Band 5, the beam FWHM is $1.36''\times0.95''$, with $\sigma_c=37$~mJy at 183 GHz (H$_2$O) and $\sigma_c=2.8$~mJy at 193~GHz (CH$_3$OH). Spectral lines were identified using frequency and intensity data from the Splatalogue \cite{rem16}. Significant line emission (identified as contiguous channels above the $2\sigma$ level, as part of a feature with a peak strength $>3\sigma$) was spectrally integrated to produce the moment 0 maps shown in Figure 1 and Extended Data Figure 2.

\subsection*{Radiative Transfer Modeling.}

The calibrated ALMA interferometric data (visibilities) were modeled in the Fourier domain using the SUBLIME 3D radiative transfer software \cite{cor22}, following a similar method to \cite{cor23}. {SUBLIME is a time-dependent, non-LTE spectral modelling code that calculates the molecular excitation, and emission and absorption of radiation in an expanding cometary atmosphere, taking into account molecular (and electronic) collisions, solar pumping, and optical depth effects.}  The Fourier domain is used since it is native to the ALMA data, and therefore intrinsically avoids the imaging artifacts that may be introduced by resampling, gridding, interpolation, and numerical deconvolution. The coma was divided into two solid-angle regions ($\Omega_1$, $\Omega_2$), each with an independent water production rate ($Q_1$, $Q_2$), outflow velocity ($v_1(r)$, $v_2(r)$) and kinetic temperature ($T_1(r)$, $T_2(r)$) as a function of radius $r$ from the nucleus. The first solid angle region ($\Omega_1$) is defined by a cone of half-opening angle $\theta_{jet}$, with its apex at the center of the nucleus, and its axial vector pointing at a phase angle $\phi$ with respect to the observer, and at a position angle $\psi$ in the plane of the sky (see Fig. 8 of ref. \cite{cor23}). The second region ($\Omega_2$) represents the remaining (ambient) coma.  Previous mm-wave spatial-spectral coma studies found that such a model geometry allows a sufficiently good fit to the observations in the presence of enhanced coma expansion velocities and outgassing rates in the (near)-sunward direction, reproducing very well the spectral line profiles (including asymmetries) for all species of interest, while keeping the number of variable model parameters to a minimum \cite{rot21,cor22,cor23}. Our model assumes a uniform (steady-state) coma production rate and expansion velocity for each solid angle region, for the period of each observing session. Although comet 12P exhibited several major outbursts in the months before perihelion \cite{fer24}, its activity level was more stable during the period of our ALMA observations (around and soon after perihelion), which justifies the assumption of a quasi-steady-state coma. 

Spectroscopic parameters for H$_2$O, HDO and CH$_3$OH were taken from the Leiden Atomic and Molecular Database LAMDA database \cite{van20}. Updated p-H$_2$O--H$_2$O collisional excitation rates were included from \cite{man24}, while HDO and CH$_3$OH collisional excitation was calculated using the thermalization approximation \cite{cro87,boc12}, with an average collisional cross-section with H$_2$O of $5\times10^{-14}$~cm$^{-2}$ \cite{biv99}. Electron collision rates were calculated using the Born approximation \cite{iti72}, following the method of \cite{biv99}, with an electron density scaling factor of $x_{ne}=0.2$ (as recommended by \cite{har10,biv19}). Radiative pumping rates were obtained from the Planetary Spectrum Generator \cite{vil18}, appropriately scaled for the comet's heliocentric distance. Molecular photodissociation rates for the active Sun were obtained from \cite{hue15}. 

To derive H$_2$O and HDO production rates, we first performed independent fits to the two CH$_3$OH datasets (April 18--25 and May 4), using models with variable $Q_1$, $Q_2$, $T_1(r)$, $T_2(r)$, $v_1$, $v_2$, $\phi$ and $\theta_{jet}$. Models were initialized using $\theta_{jet}=90^{\circ}$, $\phi=\phi_{STO}$, $\psi=$ PSAng (\emph{i.e.} representing a hemispherically-asymmetric model along the sun-comet line; $\phi_{STO}$ and PSAng are given in Supplementary Table 1).  For April 18--25, the mean $\phi_{STO}$ and PSAng values across the 9 observing periods were used ($30^{\circ}$ and $100^{\circ}$, respectively). During model optimization, the $\chi^2$ statistic was minimized across all baselines and spectral channels with the {\tt lmfit} Python code \cite{new16}, using the Levenberg-Marquardt nonlinear least-squares algorithm. On both sets of dates, the observed CH$_3$OH line-widths in the reduced image cubes were found to increase significantly as a function of distance from the nucleus, so for the CH$_3$OH models we added an acceleration term to the coma expansion velocities as a function of $r$, to obtain an improved fit to the data. For April 18--25, the data were well fit using a log-linear velocity law, increasing from $v_1=0.583\pm0.007$~km\,s$^{-1}$ at the nucleus surface ($r_n=17$~km) to $v_1=1.323\pm0.008$~km\,s$^{-1}$ at the outer edge of the model ($r_o=2\times10^5$~km), for region $\Omega_1$, and $v_2=0.540\pm0.005$~km\,s$^{-1}$ at $r_n$, increasing to $0.878\pm0.006$~km\,s$^{-1}$ at $r_o$ for region $\Omega_2$. The shapes of the corresponding best-fitting velocity profiles are shown in Supplementary Figure 1. The best-fitting $\theta_{jet}$, $\phi$ and $\psi$ values were $63.7^{\circ}\pm0.4^{\circ}$,  $30.2^{\circ}\pm0.2^{\circ}$ and $89.7^{\circ}\pm0.2^{\circ}$, respectively, indicative of a $\sim10^{\circ}$ offset between the central axis of the jet and the Sun-comet vector.  

For May 4, the higher spectral resolution of 122 kHz allowed for more detail in the fitted velocity profiles, so we explored different $v(r)$ parameterizations, informed by prior observations of coma gas acceleration \cite{lam87,tse07}, and with reference to the coma physical models of \cite{ten08,sho16} and \cite{cor21}. A two-component log-linear velocity law was found to provide a good match with the observations, with a variable inflection point at radius $r_k$, to approximate the behaviour of the multi-fluid model of \cite{sho16}. The resulting, best-fit model outflow velocities were $v_1=0.671\pm0.024$~km\,s$^{-1}$, $v_2=0.368\pm0.023$~km\,s$^{-1}$ at $r_n$, $v_1=1.042\pm0.015$~km\,s$^{-1}$, $v_2=0.840\pm0.014$~km\,s$^{-1}$ at $r_k=11,760\pm800$~km, and $v_1=2.597\pm0.412$~km\,s$^{-1}$, $v_2=2.480\pm0.482$~km\,s$^{-1}$ at $r_o$.  The shapes of the corresponding $v_1(r)$, $v_2(r)$ velocity profiles are shown in Supplementary Figure 1. The best-fitting $\theta_{jet}$, $\phi$ and $\psi$ values were $96.1^{\circ}\pm2.0^{\circ}$,  $23.8^{\circ}\pm1.2^{\circ}$ and $127.5^{\circ}\pm0.2^{\circ}$, respectively, again indicative of a $\sim10^{\circ}$ offset between the central axis of the jet and the sun-comet vector on this date.  Such an offset between the orientation of the jet and the sun-comet vector can be explained as a result of the combined effects of diurnal heating and nucleus rotation, which would tend to result in highest surface temperatures at some point in the afternoon, comet local time, depending on the nucleus rotation velocity and surface thermal inertia.

For the coma temperature retrievals, we followed the methodology of \cite{cor23}, incorporating a smoothed, segmented, linear $T(r)$ profile. We began with the simplest assumption of a constant kinetic temperature, then added complexity until a good fit to the data was obtained. This strategy kept the number of free parameters at a minimum (therefore keeping the $\chi^2$ minimization computationally feasible), while ensuring that there were enough degrees of freedom in the model to reproduce the data, but not so many that the model became ill-conditioned (manifesting as large-amplitude ripples in $T(r)$). For the April 18--25 data, we found that four segments were required in the $T(r)$ profile to obtain a good fit to the observations, whereas on May 4, only 3 segments were required. On both dates, contrary to \cite{cor23}, we found no evidence for a variable temperature profile between the sunward and anti-sunward coma regions, so the same $T(r)$ profile was used throughout the entire coma (\emph{i.e.} $T_1(r)=T_2(r)=T(r)$).  The best-fit $T(r)$ profiles on both sets of dates are shown in Extended Data Figure 3. The quality of the corresponding model fits to the spectrally and spatially resolved CH$_3$OH visibility data can be seen in Extended Data Figures 5 and 6. Uncertainties in the coma kinetic temperature profile were found to have negligible impact on our retrieved D/H value ($<3$\%).

Although accelerated outflow models were found to produce a better fit to the CH$_3$OH observations, with more trustworthy associated temperature profiles, for May 4, a single-component velocity profile model, as well as a model with constant outflow velocity, were both found to result in a similar CH$_3$OH abundance to the two-component $v(r)$ profile (all three models had best-fitting abundances within 5\% of each other).  For consistency with the H$_2$O and HDO models (for which the observational data contained insufficient information to properly constrain the coma acceleration, jet orientation and opening angle), we therefore adopted constant outflow-velocity models, with $\theta_{jet}$ fixed to $90^{\circ}$, $\phi=\phi_{STO}$ and $\psi=$ PSAng, to produce the model results shown in Table \ref{tab:results}.

Due to (1) the importance of the coma H$_2$O density as a factor in the CH$_3$OH rotational excitation, and (2) the temperature-dependence of the H$_2$O excitation, the retrieved coma kinetic temperature profile ($T(r)$) and the H$_2$O production rate ($Q({\rm H_2O}$)), are mutually inter-dependent. Therefore, for the 2024 May 4 data (when H$_2$O and CH$_3$OH were observed simultaneously), we iteratively improved $T(r)$ and $Q({\rm H_2O})$ by running successive retrievals on CH$_3$OH and H$_2$O, feeding the results of one retrieval into the other, until the respective model parameters converged. The resulting production rates and abundances are shown in Table \ref{tab:results}. As mentioned above, there was no significant evidence for deviation of H$_2$O or HDO from a simplified ``hemispherically asymmetric'' coma structure, with $\theta_{jet}=90^{\circ}$, $\phi=\phi_{STO}$, $\psi=$ PSAng, so these model parameters were held fixed, while $v_1$, $v_2$, $Q_1$ and $Q_2$ were free to vary.  {Our best-fitting SUBLIME model shows that observed H$_2$O 183~GHz line was relatively optically thin ($\tau<0.3$), making it a reliable tracer of the total H$_2$O column (in contrast to \emph{e.g.} the optically thick $1_{1,0}$--$1_{0,1}$ line at 557~GHz \cite{har11a}}.

To monitor the temporal variability of the coma CH$_3$OH/H$_2$O abundance ratio, a series of H$_2$O and CH$_3$OH observations were obtained using NASA's Infrared Telescope Facility (IRTF) on three nights, spanning the period 2024 April 18 to 2024 April 30 (see the following Section). Based on these observations, $Q({\rm CH_3OH})/Q({\rm H_2O})$ was found to show no significant evidence for variability throughout the period of our ALMA observations. Therefore, we calculated the $Q({\rm H_2O})$ value appropriate for April 18--25, based on the May 4 CH$_3$OH/H$_2$O ratio ($(2.328\pm0.016)$\%), by running a series of CH$_3$OH model fits, varying $Q({\rm H_2O})=Q_1+Q_2$ until the retrieved CH$_3$OH abundance matched the May 4 value. The required $Q({\rm H_2O})$ value of $9.23\times10^{29}$~s$^{-1}$ was then used to derive the HDO/H$_2$O mixing ratio ($x({\rm HDO})$) by fitting the HDO visibility data, with $x({\rm HDO})$, $v_1$, $v_2$ and the hemispheric asymmetry ratio ($Q_1/Q_2$) as free parameters. The best fitting models for the binned H$_2$O and HDO visibility spectra are shown in Extended Data Figure 4. Our adopted $Q({\rm H_2O})$ value for April 18--25 is further validated by its consistency with the average IRTF value of $(1.1\pm0.2)\times10^{30}$~s$^{-1}$ observed April 18 and 22. To allow for uncertainties in $Q({\rm H_2O})$ due to possible temporal variability in the $Q({\rm CH_3OH})/Q({\rm H_2O})$ ratio, an additional uncertainty of 20\% was assumed. 

Finally, we investigated the possibility of H$_2$O and HDO coma production, for example, as a result of icy grain sublimation in the coma, by performing a $\chi^2$ analysis to determine upper limits on the H$_2$O and HDO production scale lengths ($L_p$). Using a hemispherically-asymmetric Haser daughter model \cite{has57} based on the physical structure of the best-fitting parent models (Table \ref{tab:results}), the H$_2$O and HDO model parameters were re-optimized for a set of discrete $L_p$ values in the range 0--2000 km. The resulting $\chi^2$ values for each model were then plotted as a function of $L_p$, using cubic spline interpolation between the points. Upper limits on the $L_p$ values for H$_2$O and HDO (at 99\% confidence) were determined from where the respective $\chi^2(L_p)$ curves crossed a threshold of $\chi^2-\chi_m^2=6.63$, where $\chi_m^2$ is the $\chi^2$ minimum, resulting in $L_p<135$~km for H$_2$O and $L_p<610$~km for HDO. We performed a similar analysis using a composite (parent + daughter) H$_2$O model, with a ratio of parent-to-daughter production rates ($Q_p/Q_d$) fixed to a value of 1, to represent the case of equal water production from coma (daughter) and nucleus (parent) sources. The resulting upper limit for the daughter component was $L_p<237$~km. We also ran HDO models with $Q_p/Q_d$ = 1, but in this case, the $L_p$ value for the daughter component was unconstrained due to the lower signal-to-noise ratio of the HDO data.

{If H$_2$O and HDO were both produced in the coma from icy grain sublimation, in the same relative proportion as found in the nucleus-sublimated gases, then our measured D/H ratio and associated errors would not change significantly. However, in the case of HDO (or H$_2$O) being produced only from coma icy grains, while H$_2$O (or HDO) is produced directly from the nucleus, then our error bars would be underestimated. We therefore ran an additional model fit, assuming HDO is produced in the coma at $L_p=262$~km (corresponding to the $1\sigma$ uncertainty on our retrieved $L_p$(HDO) value), with H O originating only from the nucleus, and derived an additional uncertainty of 10.3\% for the HDO abundance.}

{Our total HDO/H$_2$O error budget is thus made up of the statistical uncertainties on the $Q({\rm CH_3OH})$/$Q({\rm H_2O})$ and $Q({\rm HDO})$/$Q({\rm H_2O})$ ratios (3\% and 13\%, respectively), combined in quadrature with a 20\% systematic uncertainty on $Q({\rm CH_3OH})$/$Q({\rm H_2O})$, 10\% uncertainty due to the possible range of HDO scale lengths, and an additional 4\% due to errors on the H$_2$O ortho-to-para ratio (see the following section), to produce a total ($1\sigma$) error estimate of 26\%. This leads to a final HDO/H$_2$O value of $(3.41\pm0.89)\times10^{-4}$.}

\subsection*{IRTF Observations.} High-resolution ($\Delta{\lambda}/\lambda\sim45,000$), near-infrared spectroscopic observations of 12P/Pons-Brooks were acquired using the iSHELL instrument at the NASA Infrared Telescope Facility on 2024 April 18, 22 and 30. The data were acquired using the $0.75''$ wide $\times 15''$ long slit, and the slit position angle was oriented along the projected Sun-comet line on all dates.  CH$_3$OH was sampled using the \emph{Lp1} setting, covering $\sim3.3$--3.6~$\mu$m and H$_2$O was measured using a custom L-band setting that covers 2.8--3.1~$\mu$m. The observing and data reduction followed the procedures outlined in \cite{dis17} and references therein. The comet was observed using an ABBA nodding sequence, with the A and B positions located to either side of the slit center and separated by half its length. These were combined in an A $-$ B $-$ B $+$ A sequence to isolate the comet signal. 

After flat fielding, cleaning, and straightening the curved echelle orders, spectra were extracted by summing over 9 pixel rows centered on the peak of the dust emission intensity.  Atmospheric transmittance spectra were retrieved using the NASA Planetary Spectrum Generator (PSG) \cite{vil18}, using the column burdens that were determined from a standard star spectrum obtained close in time and using the same spectral setting (though with a $4.0''$ wide slit). Subtracting these transmittance spectra, scaled to the comet continuum level, isolated cometary emission lines were then fit with synthetic fluorescence emission models (\cite{vil12b,vil12,dis13,vil18}) to determine rotational temperatures and nucleus-centered production rates. 

{Production rates were determined using the well-established $Q$-curve formalism \cite{bon05}, which specifically accounts for asymmetries in cometary gas outflow along the slit, as well as correcting for slit flux-losses near the peak of the emission profiles. Extended Data Table 1 shows the ``terminal'' (or ``global'' $Q$ values), obtained through several steps. First, production rates are measured for different rectangular apertures along the slit. For each of these a spherically symmetric outflow is assumed. However, these ``spherical'' $Q$ values are not adopted as final production rates. Instead, we calculate $Q$ by taking the mean of two production rates obtained from apertures that are equidistant from the peak intensity, on opposite sides of the long-slit spatial profile. Detailed Monte Carlo coma simulations \cite{xie96} show that averaging the $Q$ values in this way (for corresponding equidistant fields-of-view), explicitly accounts for the effects of asymmetries along the long-slit profile. The coma outflow velocity was assumed to vary as $0.8/r_H^{0.5}$, which is in agreement with the value of 0.9~km\,s$^{-1}$ measured using ALMA for the sunward hemisphere of the coma.} 

{The critical role of IR observations in this study is to demonstrate that the CH$_3$OH/H$_2$O relative abundances in the coma did not vary significantly over time between the dates of the ALMA HDO and H$_2$O observations. We therefore point out that the spatial profiles and rotational temperatures for both H$_2$O and CH$_3$OH were consistent, suggesting a common outgassing source for both molecules. Furthermore, since the observed lines are optically thin, uncertainties in the IRTF spectral model parameters (outflow velocities) and model assumptions (in the $Q$-curve formalism) affect $Q({\rm H_2O})$ and $Q({\rm CH_3OH})$ in an identical manner, making their relative abundances highly reliable.}

{The ortho-to-para ratio (OPR) of H$_2$O in comet 12P is a crucial parameter in converting the p-H$_2$O production rate (observed using ALMA) to a total $Q({\rm H_2O})$ value. The majority of OPR measurements in comets are consistent with statistical equilibrium (OPR = 3.0; \cite{che22,bon07,bon13,vil11}). However, there are exceptions \cite{fau19}.  We therefore derived the OPR in comet 12P from our IRTF spectral data using the most detailed methodology described by \cite{bon13}. By applying both global spectral fits and line-by-line analysis, the OPR retrieval takes into account stochastic as well as systematic uncertainties. Because the observed infrared emissions are optically-thin, the relative intensities between all IR lines are highly diagnostic of rotational temperature in the inner coma ($T_{rot}$), while the relative intensities between ortho- and para-lines tightly constrain the OPR. The use of a different outflow velocity, or the introduction of coma asymmetries into the spectral model, such as the hemispheric asymmetry used to analyze the ALMA data, does not impact the derived OPR. The reliability of our measurements is further improved due to the large iSHELL bandwidth, simultaneously sampling $\sim60$ ortho lines and $\sim30$ para lines. Our H$_2$O spectral models incorporated the most up-to-date (and rigorously validated) fluorescence emission models from PSG \cite{vil18}, which separately treats the effects of $T_{rot}$ and OPR on the relative line intensities.}

{A value of OPR = $3.05\pm0.18$ was derived for 2024-04-18. No evidence was found for time-variability of the OPR across the three IRTF observation dates, and all observations were found to be consistent with the statistical equilibrium value of OPR = 3.}

\subsection*{HDO Line Frequency Measurements.}

Transition frequencies for HDO reproduced in current databases date back to 1971 \cite{del71}, and the tabulated frequency of the 2$_{1,1}$--2$_{1,2}$ transition ($241561.550\pm0.037$ MHz) was found to be of insufficient accuracy for properly modeling the observed HDO line profile (where accuracies of $\lesssim10$ kHz are desirable). We therefore performed new laboratory measurements of the HDO rotational spectrum using the broadband millimeter-wave spectrometer at the Polish Academy of Sciences in Warsaw, Poland \cite{kis10}.  The sample of HDO was prepared by mixing equal quantities of H$_2$O and D$_2$O, and the resulting HDO spectral line frequencies are given in Supplementary Table 3.  The 2$_{1,1}$--2$_{1,2}$ transition was measured with particular care since it showed remnant deuterium nuclear quadrupole hyperfine structure.  Its estimated measurement uncertainty is now expected to be below 10 kHz, while for the other transitions measured, the uncertainties are estimated to be $\approx10$ kHz.


\section*{Acknowledgements} This work was supported by NASA's Planetary Science Division Internal Scientist Funding Program, through the Fundamental Laboratory Research (FLaRe) work package. E.L.G. was supported by NSF grant AST-2009910. B.P.B. was supported by NSF grant AST-2009398 and NASA SSO grant 80NSSC22K1401. K.F. was supported by JSPS KAKENHI Grant Number 20H05847. Part of this research was carried out at the Jet Propulsion Laboratory, California Institute of Technology, under a contract with the National Aeronautics and Space Administration (80NM0018D0004). D.C.L. acknowledges financial support from the National Aeronautics and Space Administration (NASA) Astrophysics Data Analysis Program (ADAP). ALMA is a partnership of ESO, NSF (USA), NINS (Japan), NRC (Canada), NSC and ASIAA (Taiwan) and KASI (Republic of Korea), in cooperation with the Republic of Chile. The JAO is operated by ESO, AUI/NRAO and NAOJ.  The NRAO is a facility of the National Science Foundation operated under cooperative agreement by Associated Universities, Inc. We thank Michael R. Combi for sharing 12P water production rates derived from SOHO/SWAN observations, and for helpful discussions on the radial dependence of coma outflow velocities. B.P.B., E.L.G. and N.X.R. were Visiting Astronomers at the IRTF, which is operated by the University of Hawaii under contract NNH14CK55B with the National Aeronautics and Space Administration. The authors wish to recognize and acknowledge the very significant cultural role and reverence that the summit of Maunakea has always had within the indigenous Hawaiian community. We are most fortunate and grateful to have the opportunity to conduct IRTF observations from this mountain.

\setcounter{figure}{0}
\renewcommand{\figurename}{Extended Data Figure}

\begin{figure*}
\centering
\includegraphics[width=\textwidth]{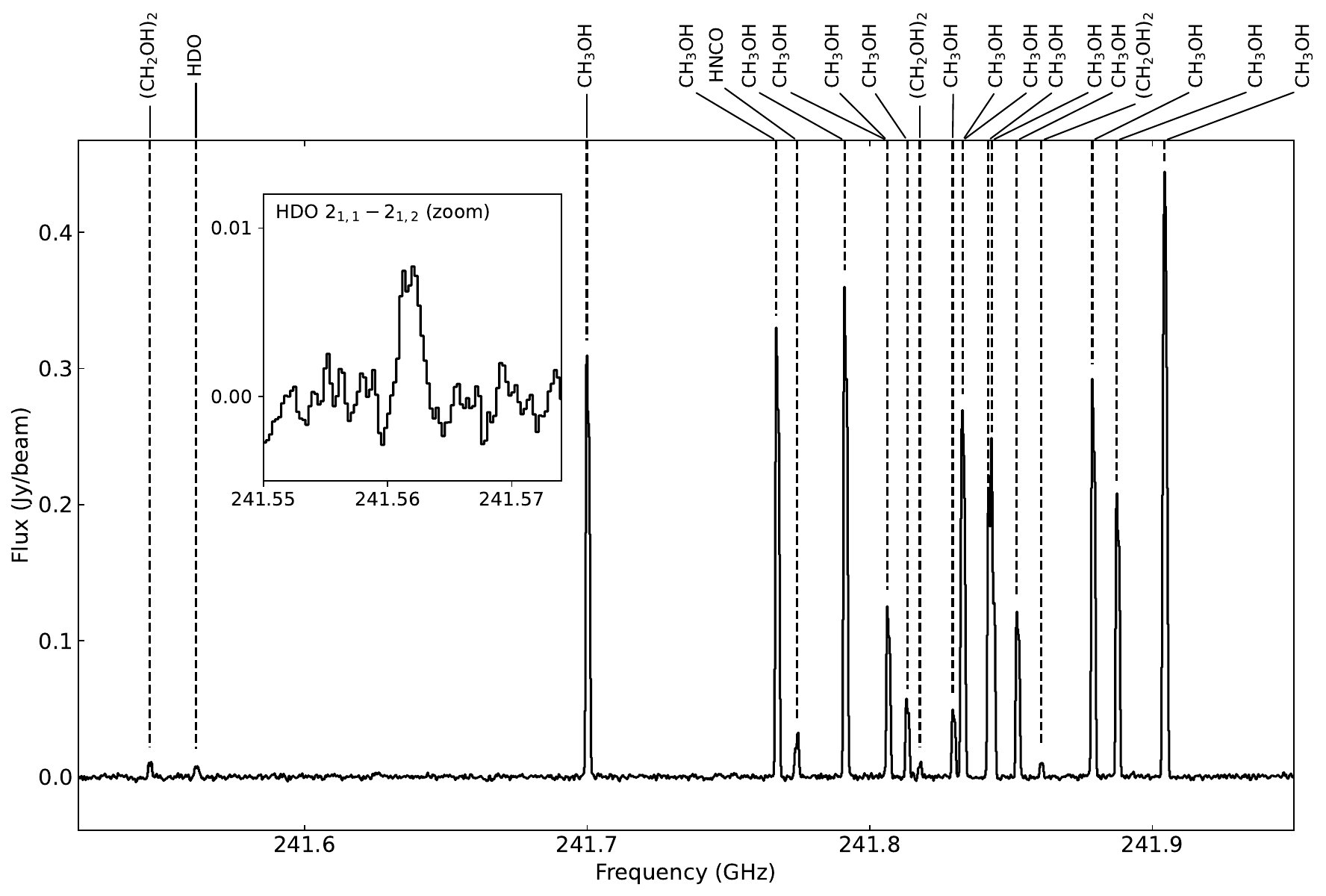}
\caption{{\bf Observed ALMA spectrum of comet 12P.} These data were observed at the continuum emission peak (the assumed location of the nucleus) using ALMA on 2024 April 18--25, in the vicinity of the HDO $2_{1,1}$--$2_{1,2}$ and CH$_3$OH $J_K=5_K$--$4_K$ lines (labeled). The spectrum has been continuum-subtracted, and molecular lines were assigned using data from the Splatalogue (https://splatalogue.online/) \cite{rem16}. Inset panel shows a zoom of the detected HDO line.
\label{fig:spec}}
\end{figure*}

\begin{figure*}
\centering
\includegraphics[width=\textwidth]{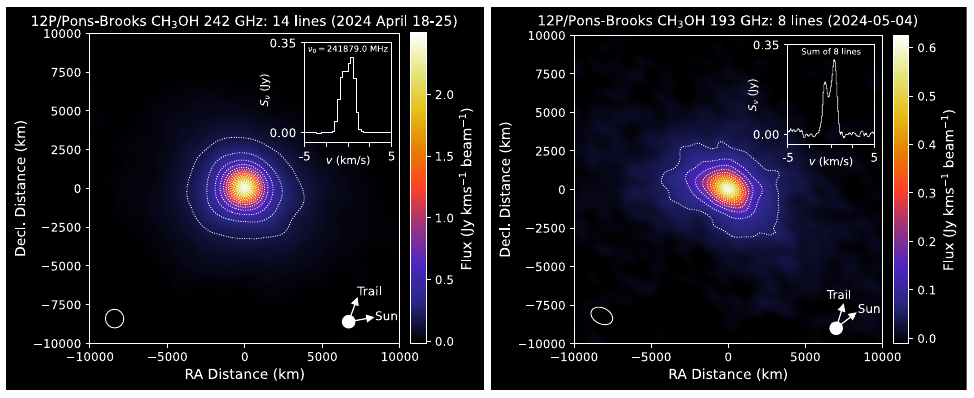}
\caption{{\bf ALMA emission maps of CH$_3$OH in comet 12P}. The data, observed in Band 6 (between 2024 April 18--25; left) and Band 5 (2024 May 4; right), have been spectrally integrated, continuum-subtracted, and are centered on the thermal continuum peak. The Band 6 image is integrated over 14 transitions, with a contour separation of $10\sigma$, while the Band 5 image is integrated over 8 transitions, with $5\sigma$ contours (see Supplementary Table 2 for details). Inset panels (upper right) show the spectral line profiles of selected transitions as a function of cometocentric velocity ($v$), extracted at the origin (8 lines in the Band 5 CH$_3$OH spectrum have been averaged in velocity space, to improve the signal-to-noise rato). The spatial resolution (beam FWHM) is shown lower left, and the directions of the (sky-projected) comet-Sun and orbital trail vectors are shown lower right.
\label{fig:ch3ohmaps}}
\end{figure*}

\begin{figure*}
\centering
\includegraphics[width=0.7\textwidth]{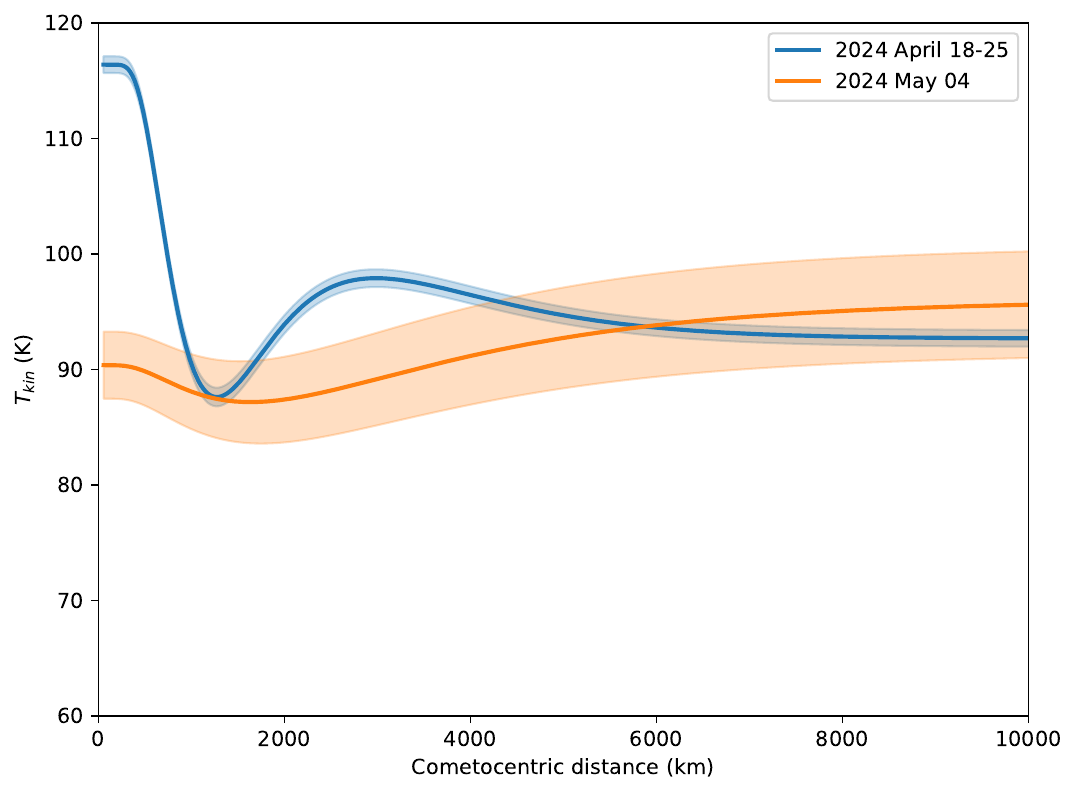}
\caption{{\bf Coma kinetic temperature radial profiles.} Temperatures were retrieved independently on the HDO and H$_2$O observation dates dates using multi-line ALMA CH$_3$OH observations. $1\sigma$ error envelopes are shown.
\label{fig:temp}} 
\end{figure*}

\begin{figure*}
\centering
\includegraphics[width=\textwidth]{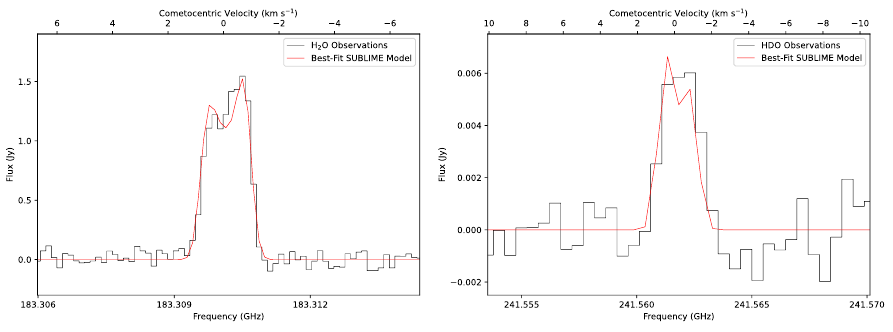}
\caption{{\bf ALMA interferometric spectra of H$_2$O and HDO.} The real part of the observed visibility data is shown as a function of frequency for H$_2$O (left) and HDO (right), averaged across the antenna baseline range $uv=12$--100 m (corresponding to angular scales on the sky $<24''$ for H$_2$O and $<21''$ for HDO). Best-fit SUBLIME radiative transfer models (see Table \ref{tab:results} and Figure \ref{fig:vis} for details), are overlaid in red.
\label{fig:specfits}}
\end{figure*}

\begin{figure*}
\centering
\includegraphics[width=\textwidth]{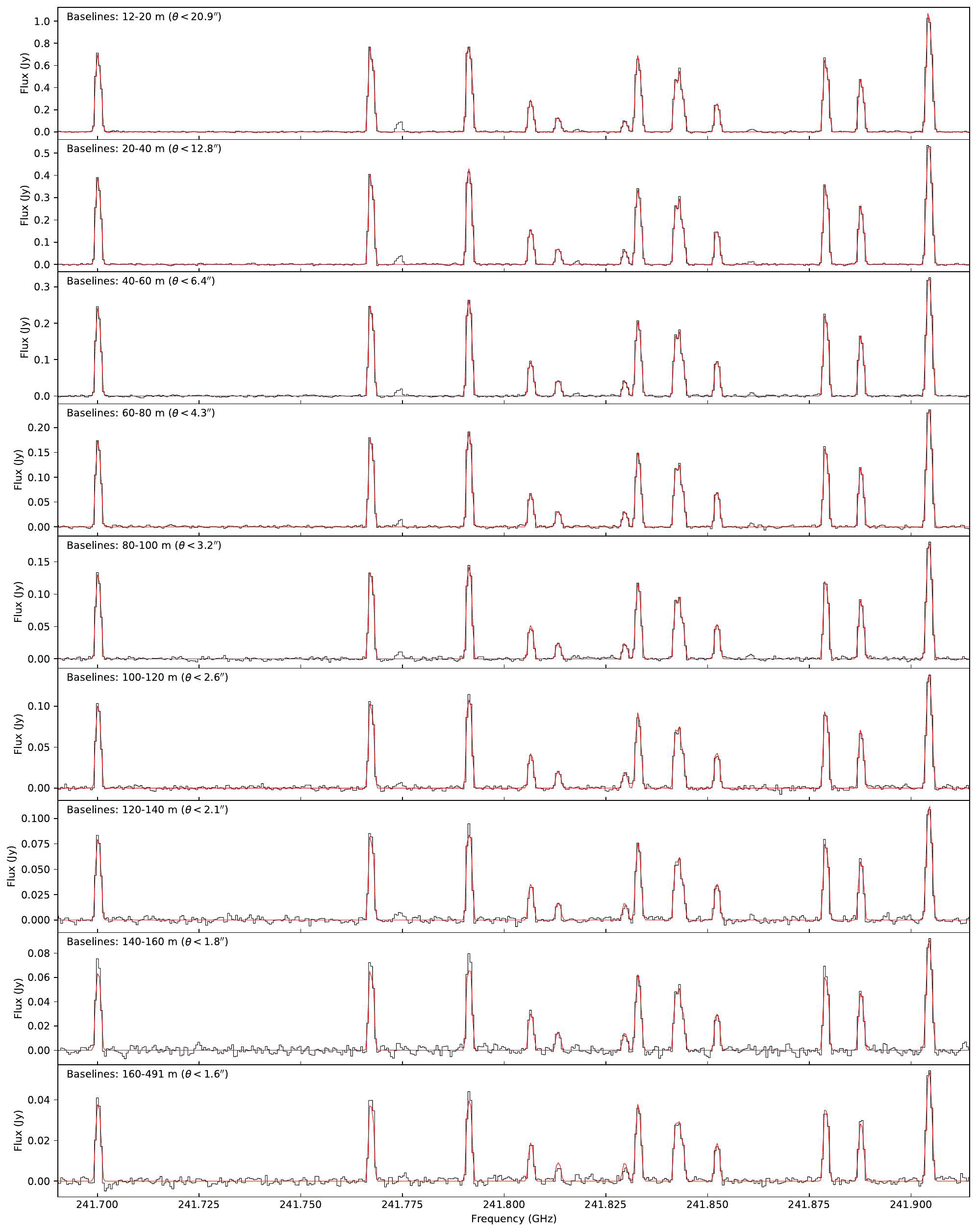}
\caption{{\bf ALMA interferometric spectra of CH$_3$OH (average of observations in the date range 2024 April 18--25).} The real parts of the observed visibility spectra are shown as a function of baseline interval, averaged within the $uv$ range specified in each panel (corresponding to angular sizes $\theta$ in the plane of the sky). The best-fit SUBLIME radiative transfer model is overlaid in red.
\label{fig:ch3ohspecfits1}}
\end{figure*}

\begin{figure*}
\centering
\includegraphics[width=\textwidth]{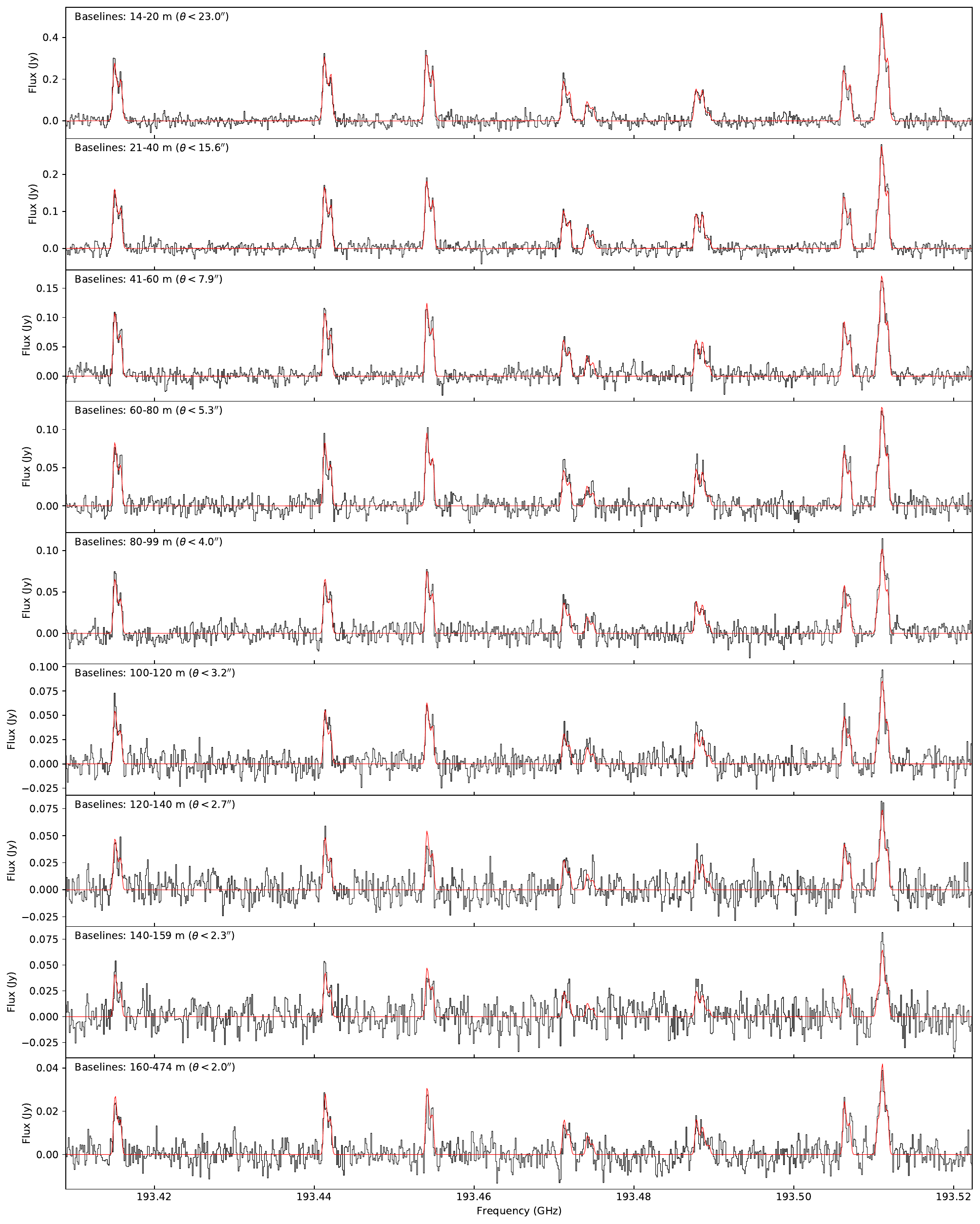}
\caption{{\bf ALMA interferometric spectra of CH$_3$OH (observed 2024 May 4).} The real parts of the observed visibility spectra are shown as a function of baseline interval, averaged within the $uv$ range specified in each panel (corresponding to angular sizes $\theta$ in the plane of the sky). The best-fit SUBLIME radiative transfer model is overlaid in red.
\label{fig:ch3ohspecfits2}}
\end{figure*}

\setcounter{table}{0}
\captionsetup[table]{name=Extended Data Table}

\begin{table*}
\centering
\caption{IRTF Measurements\label{tab:irtf}}
\begin{tabular}{cllll}
\hline\hline
Date & Molecule & $T_{rot}$ & $Q$ & $Q({\rm CH_3OH})/Q({\rm H_2O})$ \\
     &          & (K)       & ($10^{28}$~s$^{-1}$)     & (\%)\\
\hline  
2024-04-18& H$_2$O &  $98\pm2$   & $94\pm9$    & $1.7\pm0.3$\\
          & CH$_3$OH &             & $1.6\pm2$   &\\
2024-04-22&  H$_2$O &  $102\pm2$  & $120\pm20$  & $1.8\pm0.3$\\
          & CH$_3$OH &             & $2.2\pm0.1$ &\\
2024-04-30& H$_2$O  &  $100\pm2$  & $83\pm5$    & $1.6\pm0.4$\\
          & CH$_3$OH &             & $1.3\pm0.2$  &\\
\hline
\end{tabular}

\end{table*}

\clearpage



\section*{Supplementary material}

# Supplementary Information

## A D/H Ratio Consistent with Earth's Water in Halley-type Comet 12P from ALMA HDO Mapping

M. A. Cordiner et al.

## Supplementary Tables

Supplementary Table 1: Observational Data

| Date (UT) | Obs. Time (UT) | Freqs. (GHz) | PWV[a] (mm) | Ants.[b] | Baselines (m) | $\Delta$[c] (au) | $r_H$[d] (au) | $\epsilon$[e] (°) | $\phi_{STO}$[f] (°) | PSAng[g] (°) | PSAMV[h] (°) |
|---|---|---|---|---|---|---|---|---|---|---|---|
| 2024-04-18 | 17:58–18:59 | 242-259 | 1.00 | 43 | 15-440 | 1.61 | 0.78 | 22.3 | 29.6 | 91.2 | 338.3 |
| 2024-04-19 | 16:31–17:41 | 242-259 | 0.56 | 44 | 15-456 | 1.61 | 0.78 | 23.0 | 29.6 | 93.6 | 338.7 |
| 2024-04-19 | 17:50–19:00 | 242-259 | 0.63 | 43 | 15-456 | 1.61 | 0.78 | 22.3 | 29.6 | 93.7 | 338.8 |
| 2024-04-19 | 19:16–20:26 | 242-259 | 0.76 | 42 | 15-456 | 1.61 | 0.78 | 22.5 | 29.7 | 94.0 | 338.8 |
| 2024-04-23 | 17:09–18:21 | 242-259 | 1.10 | 42 | 15-484 | 1.60 | 0.78 | 23.6 | 30.1 | 103.9 | 340.6 |
| 2024-04-23 | 18:24–19:36 | 242-259 | 1.19 | 42 | 15-484 | 1.60 | 0.78 | 23.7 | 30.1 | 103.9 | 340.6 |
| 2024-04-24 | 17:56–19:08 | 242-259 | 1.22 | 43 | 15-484 | 1.60 | 0.78 | 23.8 | 30.2 | 106.4 | 350.0 |
| 2024-04-24 | 19:09–20:21 | 242-259 | 0.97 | 43 | 15-484 | 1.60 | 0.78 | 23.8 | 30.2 | 106.5 | 350.0 |
| 2024-04-25 | 18:40–19:50 | 242-259 | 1.76 | 46 | 15-500 | 1.60 | 0.79 | 23.9 | 30.5 | 109.0 | 341.4 |
| 2024-05-04 | 15:43–16:29 | 181-194 | 0.39 | 41 | 15-500 | 1.59 | 0.82 | 27.4 | 33.4 | 128.0 | 343.7 |

Table Footnotes — [a]Precipitable water vapour at zenith; [b]Number of active ALMA antennas; [c]Geocentric distance; [e]Solar elongation angle; [f]Sun-target-observer (phase) angle; [g]Position angle of the extended Sun-comet vector; [h]Negative of the comet's heliocentric velocity vector, counterclockwise from north in the plane of the sky.

1

Supplementary Table 2: Molecular Line Data

| Molecule | Transition | Frequency (MHz) | $E_u$ (K) |
|---|---|---|---|
| HDO | $2_{1,1}$–$2_{1,2}$ | 241561.550 | 95 |
| CH$_3$OH | $5_0$–$4_0\ E$ | 241700.159 | 48 |
| CH$_3$OH | $5_{-1}$–$4_{-1}\ E$ | 241767.234 | 40 |
| CH$_3$OH | $5_0$–$4_0\ A^+$ | 241791.352 | 35 |
| CH$_3$OH | $5_4$–$4_4\ A^\pm$ | 241806.524 | 115 |
| CH$_3$OH | $5_{-4}$–$4_{-4}\ E$ | 241813.255 | 123 |
| CH$_3$OH | $5_3$–$4_3\ A^\pm$ | 241833.106 | 85 |
| CH$_3$OH | $5_2$–$4_2\ A^-$ | 241842.284 | 73 |
| CH$_3$OH | $5_3$–$4_3\ E$ | 241843.604 | 83 |
| CH$_3$OH | $5_{-3}$–$4_{-3}\ E$ | 241852.299 | 98 |
| CH$_3$OH | $5_1$–$4_1\ E$ | 241879.025 | 56 |
| CH$_3$OH | $5_2$–$4_2\ A^+$ | 241887.674 | 73 |
| CH$_3$OH | $5_{-2}$–$4_{-2}\ E$ | 241904.147 | 61 |
| CH$_3$OH | $5_2$–$4_2\ E$ | 241904.643 | 57 |
| p-H$_2$O | $3_{1,3}$–$2_{2,0}$ | 183310.087 | 205 |
| CH$_3$OH | $4_0$–$3_0\ E$ | 193415.324 | 36 |
| CH$_3$OH | $4_{-1}$–$3_{-1}\ E$ | 193441.600 | 29 |
| CH$_3$OH | $4_0$–$3_0\ A^+$ | 193454.358 | 23 |
| CH$_3$OH | $4_3$–$3_3\ A^+$ | 193471.434 | 73 |
| CH$_3$OH | $4_{-3}$–$4_{-3}\ A^-$ | 193471.545 | 73 |
| CH$_3$OH | $4_3$–$3_3\ E$ | 193474.414 | 71 |
| CH$_3$OH | $4_{-2}$–$3_{-2}\ A-$ | 193488.047 | 61 |
| CH$_3$OH | $4_{-3}$–$3_{-3}\ E$ | 193488.964 | 86 |
| CH$_3$OH | $4_1$–$3_1\ E$ | 193506.559 | 44 |
| CH$_3$OH | $4_2$–$3_2\ A^+$ | 193510.750 | 61 |
| CH$_3$OH | $4_{-2}$–$3_{-2}\ E$ | 193511.225 | 49 |
| CH$_3$OH | $4_2$–$3_2\ E$ | 193511.335 | 46 |

Table Footnotes — Transition frequencies and upper-state energy levels ($E_u$) are from the JPL Molecular Spectroscopy database (Pickett, H. M. et al. 1998, JQSRT, 60, 883).

Supplementary Table 3: New HDO Line Frequency Measurements

| Transition | Catalogue[a] (MHz) | Measured[b] (MHz) | $\Delta(f)^c$ (MHz) |
|---|---|---|---|
| $2_{1,1} - 2_{1,2}$ | 241561.550(37) | 241561.641(10) | 0.0912 |
| $7_{3,4} - 6_{4,3}$ | 241973.570(39) | 241973.595(10) | 0.0252 |
| $5_{2,3} - 4_{3,2}$ | 255050.260(59) | 255050.310(10) | 0.0499 |
| $7_{5,3} - 8_{4,4}$ | 258223.760(104) | 258223.933(10) | 0.1728 |
| $2_{2,0} - 3_{1,3}$ | 266161.070(25) | 266161.097(10) | 0.0269 |
| $9_{4,5} - 8_{5,4}$ | 272907.540(26) | 272907.536(10) | -0.0038 |
| $7_{5,2} - 8_{4,5}$ | 283318.590(23) | 283318.567(10) | -0.0234 |
| $8_{3,5} - 8_{3,6}$ | 305038.550(112) | 305038.594(10) | 0.0438 |
| $5_{2,3} - 5_{2,4}$ | 310533.290(51) | 310533.409(10) | 0.1190 |
| $6_{2,5} - 5_{3,2}$ | 313750.620(31) | 313750.681(10) | 0.0613 |
| $5_{4,2} - 6_{3,3}$ | 317151.250(48) | 317151.191(10) | -0.0594 |

Table Footnotes — [a]Catalogue HDO line frequencies (from Messer, F. C. et al. 1984, J. Mol. Spectrosc. 105, 139). [b]Measured frequencies (this work). [c]Difference ($b - a$).

# Supplementary Figures

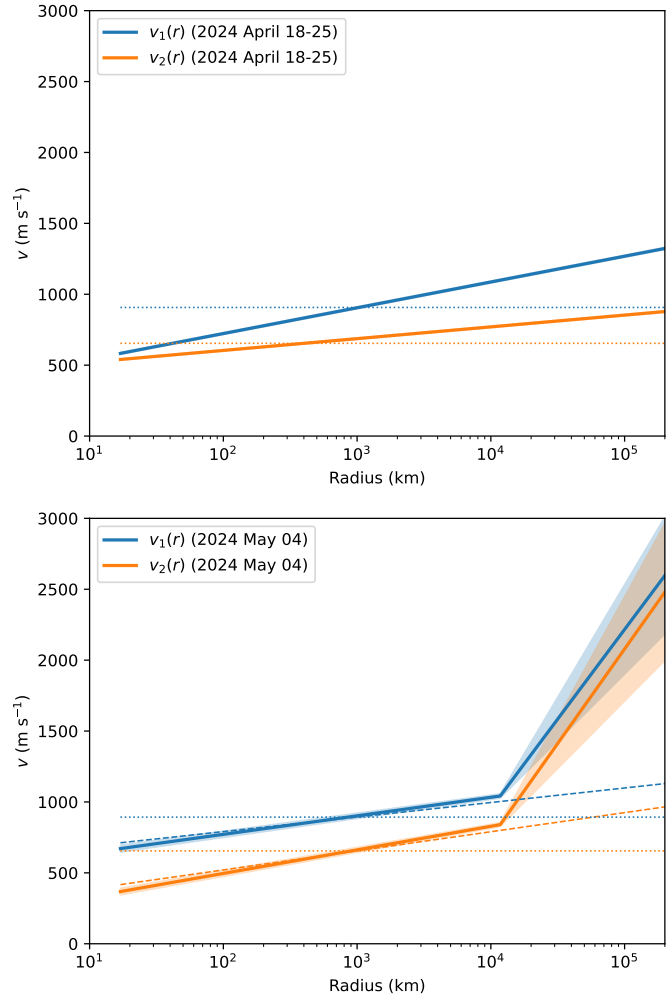

Supplementary Figure 1: Retrieved coma radial expansion velocity profiles ($v(r)$) and $1\sigma$ error envelopes, based on fits to the ALMA $CH_3OH$ data on April 18–25 (top) and May 4 (bottom). Solid lines represent the best-fitting $v(r)$ profiles obtained during the coma temperature retrievals, for the sunward ($v_1(r)$) and anti-sunward ($v_2(r)$) sides of the comet (see Methods for details). Dotted lines represent the (constant) expansion velocities retrieved using a model with zero outflow acceleration. For 2024 May 04, the $v(r)$ profiles retrieved using a single-component log-linear form are also shown for reference (dashed lines).



\end{document}